\documentclass[aps,prl,twocolumn,nofootinbib,superscriptaddress]{revtex4} 

\usepackage{epsfig}
\usepackage{color}
\usepackage[latin1]{inputenc}
\usepackage{float,amsmath}
\usepackage{graphicx}

\begin{document}

\title{Moving forward to constrain the shear viscosity of QCD matter}
\author{Gabriel Denicol}
\affiliation{Physics Department, Brookhaven National Laboratory, Upton, NY 11973, USA}

\author{Akihiko Monnai}
\affiliation{RIKEN BNL Research Center, Brookhaven National Laboratory, Upton, NY 11973, USA}

\author{Bj\"orn Schenke}
\affiliation{Physics Department, Brookhaven National Laboratory, Upton, NY 11973, USA}

\begin{abstract}
We demonstrate that measurements of rapidity differential anisotropic flow in heavy ion collisions can constrain the temperature dependence of the shear viscosity to entropy density ratio $\eta/s$ of QCD matter. Comparing results from hydrodynamic calculations with experimental data from RHIC, we find evidence for a small $\eta/s\approx 0.04$ in the QCD cross-over region and a strong temperature dependence in the hadronic phase. A temperature independent $\eta/s$ is disfavored by the data. We further show that measurements of the event-by-event flow as a function of rapidity can be used to independently constrain the initial state fluctuations in three dimensions and the temperature dependent transport properties of QCD matter.
\end{abstract}

\maketitle


{\bf Introduction}
The matter produced in ultra-relativistic heavy ion collisions at the Relativistic Heavy Ion Collider (RHIC) and the Large Hadron Collider (LHC) has been shown to behave like an almost perfect fluid. It is well described by viscous relativistic hydrodynamics with one of the smallest shear viscosity to entropy density ratios, $\eta/s$, ever observed (see \cite{Heinz:2013th,Gale:2013da,deSouza:2015ena} for recent reviews). So far, most hydrodynamic simulations of heavy ion collisions assume a temperature independent $\eta/s$, which is then extracted from measurements. However, it is well known that the $\eta/s$ of quantum chromodynamic (QCD) matter cannot be constant \cite{Prakash:1993bt,Arnold:2003zc}: it is expected to display a strong temperature dependence and have a minimum around the phase transition/cross-over region -- a behavior shared by many fluids in nature \cite{Csernai:2006zz}. Understanding and quantifying this temperature dependence around the transition from hadronic matter to the quark-gluon plasma (QGP) is of fundamental importance as it will reveal the true transport properties of QCD matter in the strong coupling regime.

Recent progress in the experimental precision and the study of new observables \cite{Bilandzic:2013kga,Aad:2014fla,Niemi:2015qia} has opened up the path towards a quantitative determination of the transport properties of fundamental QCD matter, in particular, the extraction of the temperature-dependence of the shear viscosity \cite{Niemi:2015qia} and even bulk viscosity \cite{Rose:2014fba,Ryu:2015vwa}. At this point, most of the theoretical effort in this direction used simplified dynamical descriptions of the collision that simulate the evolution of the produced QCD matter only in the mid-rapidity region and neglect the dynamics and fluctuations in the longitudinal direction (along the beam line).   

However, after initial state fluctuations in the transverse plane of the collision were discovered to be essential for the understanding of all observed multi-particle correlations \cite{Mishra:2007tw,Takahashi:2009na,Alver:2010gr,Alver:2010dn,Holopainen:2010gz,Schenke:2010rr,Qiu:2011hf,Gale:2012rq,Niemi:2015qia}, one must also take into account fluctuations in the longitudinal direction which can be of comparable importance \cite{Pang:2014pxa}.
With the advent of 3+1 dimensional event-by-event relativistic viscous fluid dynamic simulations \cite{Schenke:2010rr,Bozek:2011ua,Molnar:2014zha,Karpenko:2015xea}, this now becomes possible and we have theoretical access to the entire space time evolution of heavy ion collisions. This can be of particular importance to the extraction of transport coefficients since temperature (and baryon chemical potential) profiles of the medium vary in the longitudinal direction, such that particles produced with different momentum rapidities provide access to a range of varying medium properties, even at a fixed collision energy.

In this letter we propose to make use of this fact to extract the temperature dependence of $\eta/s$ from the rapidity dependence of 
experimental observables.
We employ a hydrodynamic simulation with an initial state that describes fluctuations of both net-baryon and entropy density in all three spatial dimensions.
We show that the rapidity dependence of the flow harmonic coefficients $v_2$ and $v_3$, which measure the azimuthal momentum anisotropy of the particles produced in the collision, is sensitive to the temperature dependence of $\eta/s$.
We find that agreement with experimental data requires a strong temperature dependence of $\eta/s$ at lower temperatures and a minimum value in the transition region that is considerably smaller than previous predictions made assuming a constant $\eta/s$. We also constrain the rate at which this transport coefficient can grow as the temperature becomes larger.   

We note that previous calculations within 3+1D hydrodynamics have generally not been able to describe the pseudo-rapidity dependence of $v_2$ \cite{Nonaka:2006yn,Bozek:2011ua,Molnar:2014zha}. Our results indicate that this is due to the choice of the transport parameters and their temperature dependence in these works. 

We further propose the measurement of the event-by-event distributions of the $v_n$ as functions of rapidity to constrain the three-dimensional fluctuating initial state.

{\bf Initial state model and hydrodynamic evolution}
The longitudinal fluctuations are introduced via a simple model that is a straight forward extension to the Monte Carlo Glauber model \cite{Miller:2007ri}. 
In this model, nucleons are sampled from Woods-Saxon distributions, and constituent quarks from an exponential distribution around the center of each nucleon. 
The quarks' longitudinal momentum fractions $x$ are sampled from CT10 NNLO parton distribution functions \cite{Gao:2013xoa} at $Q^2=1\,{\rm GeV^2}$ with EPS09 nuclear correction \cite{Eskola:2009uj} using LHAPDF 6.1.4 \cite{Buckley:2014ana}. Their initial rapidities are then given by $y_q = \pm y_{\rm beam} \mp \ln(1/x)$, where $y_{\rm beam}$ is the beam rapidity and the sign depends on whether the nucleus is right or left moving.
According to a sampled impact parameter, two nuclei are then overlayed and wounded quarks determined using the quark-quark cross section $\sigma_{qq}$. We use Gaussian wounding \cite{Bialas:2006qf,Broniowski:2007nz} and $\sigma_{qq}=9\,{\rm mb}$ for $\sqrt{s}=200\,{\rm GeV}$ collisions, which reproduces the nucleon-nucleon cross section of $42\,{\rm mb}$.

The distribution of quarks in rapidity after the collision is determined using a Monte Carlo implementation of the Lexus model \cite{Jeon:1997bp,Monnai:2015sca}, where the probability for a quark with rapidity $y_P$ to obtain rapidity $y$ after collision with a quark of rapidity $y_T$ (from the other nucleus) is
\begin{align}\label{eq:Q}
  Q(y-y_T,&y_P-y_T,y-y_P) = \nonumber\\
&\lambda \frac{\cosh(y-y_T)}{\sinh(y_P-y_T)} + (1-\lambda) \delta(y-y_P)\,.
\end{align}
The parameter $\lambda$ controls the degree of baryon stopping. In this work we use $\lambda=0.22$, which reproduces the experimental net-baryon distribution in Au+Au collisions at $\sqrt{s}=200\,{\rm GeV}$.
While each quark-quark collision changes both quarks' rapidity according to (\ref{eq:Q}), an entropy density is deposited between the two quarks only for the last\footnote{Ordering of collisions is done using the quarks' positions in the direction parallel to the beam line} quark-quark collision. This method leads to number of quark participant scaling of the multiplicity. Entropy density is deposited in ``tubes'' around the center of mass of the two colliding quarks and assumed to be constant in rapidity for each tube.
 The normalization 
of the entropy density for each tube is varied using negative binomial fluctuations with the parameters adjusted to reproduce the measured multiplicity distribution.\footnote{This method is only approximate because the experimental multiplicity distribution is uncorrected.}
In the transverse plane we smear the entropy density around the center of mass position of each pair by a Gaussian of width $\sigma_T=0.2\,{\rm fm}$.

This model provides fluctuating entropy and baryon density profiles that are used as initial conditions for the hydrodynamic simulation \textsc{Music} \cite{Schenke:2010nt,Schenke:2010rr,Schenke:2011bn,Gale:2012rq}. 
We use exactly the same setup as described in \cite{Monnai:2015sca}, except that we employ the relaxation time approximation to compute both bulk and shear non-equilibrium corrections to the particle distribution functions.

The equation of state at finite baryon chemical potential is constructed by interpolating the pressures of hadronic resonance gas and lattice QCD \cite{Borsanyi:2013bia,Borsanyi:2011sw} at the connecting temperature $T_c(\mu_B) =  0.166\,{\rm GeV} - 0.4(0.139\,{\rm GeV}^{-1} \mu_B^2 + 0.053\,{\rm GeV}^{-3} \mu_B^4)\,$. This ansatz is motivated by the chemical freeze-out curve determined in \cite{Cleymans:2005xv}. The temperature region below $T_c$ can be interpreted as the hadronic phase and the region above it as the QGP phase.

The initial time for the hydrodynamic evolution is $\tau_0=0.38\,{\rm fm}/c$ and kinetic freeze-out occurs at an energy density of $0.1\,{\rm GeV}/{\rm fm}^3$.

{\bf Temperature dependent transport parameters}
Similar to the investigations in \cite{Niemi:2011ix,Niemi:2012ry} and \cite{Niemi:2015qia}, we employ a simple parametrization of the temperature dependent shear viscosity to entropy density ratio $(\eta/s)(T)$. Because we allow for finite baryon chemical potential $\mu_B$ the more natural quantity to specify is $(\eta T/(\varepsilon+P))(T)$ \cite{Liao:2009gb}. At $\mu_B=0$ this equals $(\eta/s)(T)$. For most rapidities in $\sqrt{s}=200\,{\rm GeV}$ collisions $\mu_B$ is negligible and we will use $\eta T/(\varepsilon+P)$ and $\eta/s$ interchangeably in this work.

We assume a minimum at $T_c(\mu_B)$ and linear temperature dependencies above and below that minimum
\begin{align}
(\eta T/(\varepsilon+P))(T) &= (\eta T/(\varepsilon+P))_{\rm min} \notag\\ &~~~~+ a \times (T_c-T)\theta(T_c-T) \notag\\ 
& ~~~~+ b \times (T-T_c)\theta(T-T_c)\,,
\end{align}
where $a$ and $b$ are the slope parameters to be varied in the presented analysis.

\begin{figure}[htb]
\includegraphics[width=0.45\textwidth]{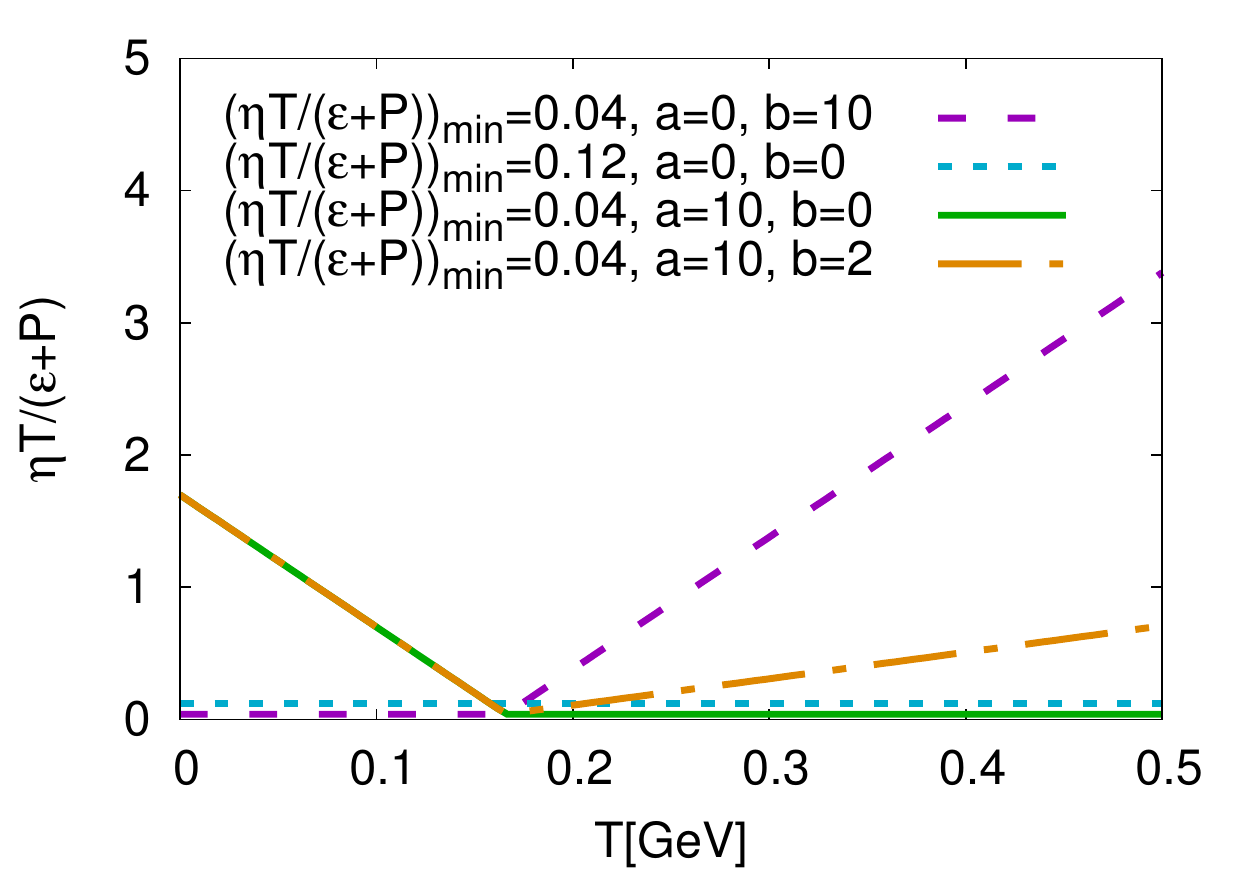}
\caption{(Color online) The four scenarios of temperature dependent $\eta T/(\varepsilon+P)$ at $\mu_B=0$. \label{fig:etaOverS}}
\end{figure}

We will study four scenarios. A constant transport parameter $\eta T/(\varepsilon+P) =0.12$, a large shear viscosity in the hadronic phase with $(\eta T/(\varepsilon+P))_{\rm min} = 0.04$, $a=10$ and $b=0$, a large viscosity in the QGP phase using $(\eta T/(\varepsilon+P))_{\rm min} = 0.04$, $a=0$ and $b=10$, and a large hadronic and moderate QGP viscosity using $(\eta T/(\varepsilon+P))_{\rm min} = 0.04$, $a=10$ and $b=2$. Figure \ref{fig:etaOverS} shows a comparison of $(\eta T/(\varepsilon+P))(T)$ in these four scenarios.

In all scenarios the shape of the bulk viscosity's temperature dependence is the same as employed in \cite{Ryu:2015vwa}, where it is assumed to peak in the transition region.
In this work the peak position is chosen to be at $T_c(\mu_B)$ and we replace the entropy density $s$ by $(\varepsilon+P)/T$ to account for the finite baryon chemical potential. Note that the inclusion of bulk viscosity has been shown to be necessary to describe the mean transverse momentum of hadrons observed at the LHC for IP-Glasma initial conditions  \cite{Ryu:2015vwa}. We remark that the same conclusion holds for the initial state used in this letter.

\begin{figure}[htb]
\includegraphics[width=0.48\textwidth]{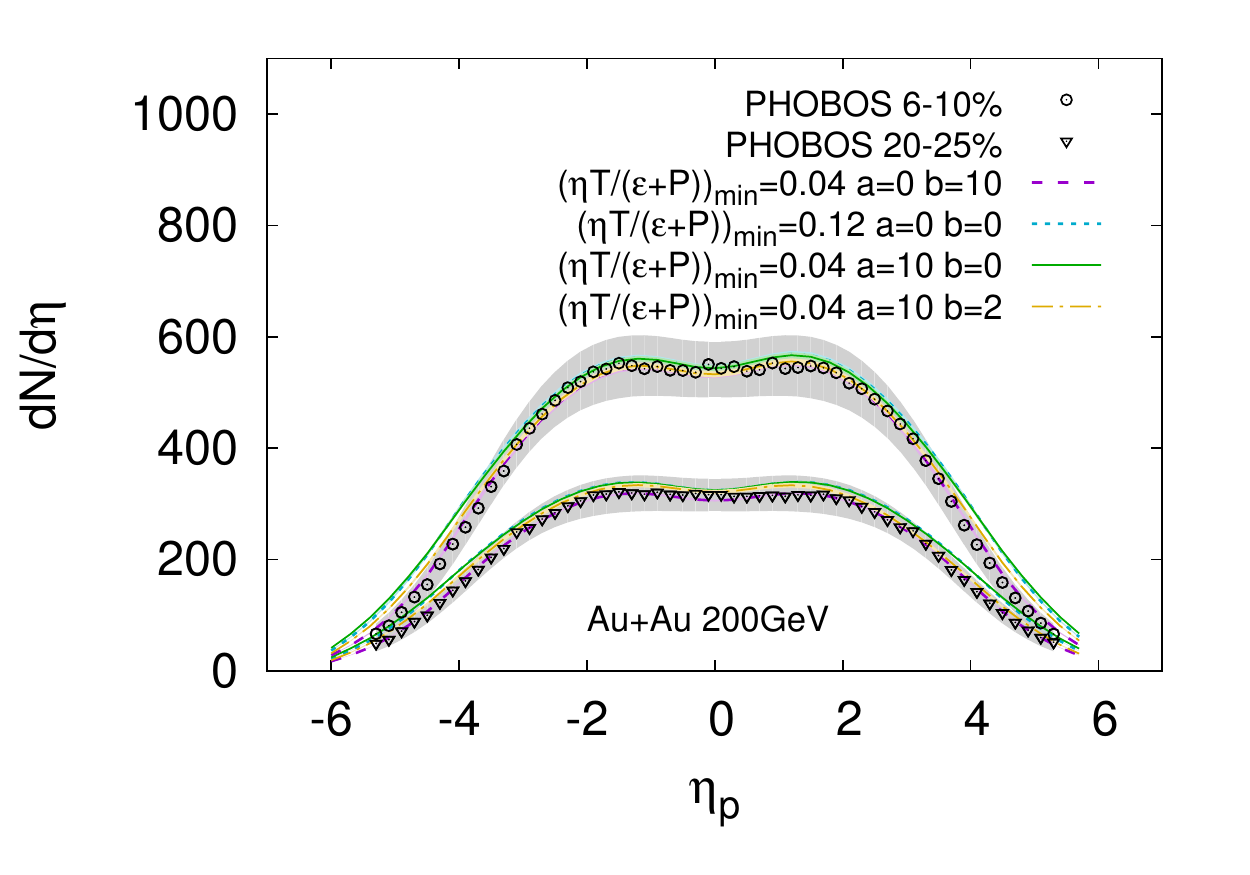}
\caption{(Color online) $dN/d\eta_p$ of charged hadrons in two different centrality classes for the four scenarios compared to experimental data from the PHOBOS collaboration \cite{Alver:2010ck}. \label{fig:dNdeta-PHOBOS}}
\end{figure}

{\bf Rapidity spectra}
We present as a baseline the results for the pseudo-rapidity dependent particle spectra in comparison to PHOBOS data \cite{Alver:2010ck} in Fig.\,\ref{fig:dNdeta-PHOBOS}. The normalization of the initial entropy density was adjusted in each scenario to fit the most central (0-3\% central) events. A large viscosity at higher temperatures inhibits the longitudinal expansion most and leads to the best description of the spectra with the used initial state model. At $\eta_p=4$, $dN/d\eta_p$ is over-estimated by approximatelty 15\% in the two scenarios with the smallest QGP viscosity.

{\bf Rapidity dependent anisotropic flow}
The flow harmonics $v_n$ as functions of pseudo-rapidity are calculated using the event average
\begin{align}\label{eq:vn}
v_n\{2\}(\eta_p) = \frac{\langle v_n v_n(\eta_p) \cos[n(\psi_n-\psi_n(\eta_p))]\rangle}{\sqrt{\langle v_n^2 \rangle}}\,.
\end{align}
 $\psi_n(\eta_p)$ is the event plane at pseudo-rapidity $\eta_p$, and $v_n$ and $\psi_n$ are the average values over the pseudo-rapidity range $|\eta_p|<6$.
We have verified that in the simulation the resulting $v_n\{2\}(\eta_p)$ are very close to the root mean square values $\sqrt{\langle v_n^2(\eta_p) \rangle}$. For clarity of notation in the following we will refer to $v_n\{2\}(\eta_p)$ from (\ref{eq:vn}) as $v_n(\eta_p)$.

\begin{figure}[htb]
\includegraphics[width=0.45\textwidth]{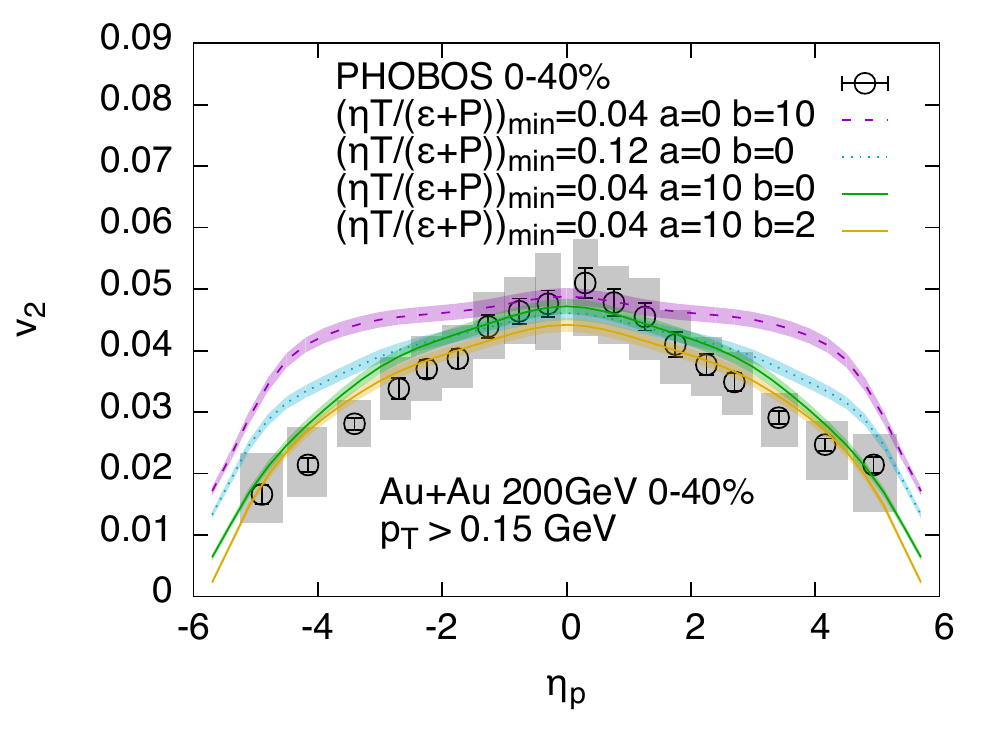}
\includegraphics[width=0.45\textwidth]{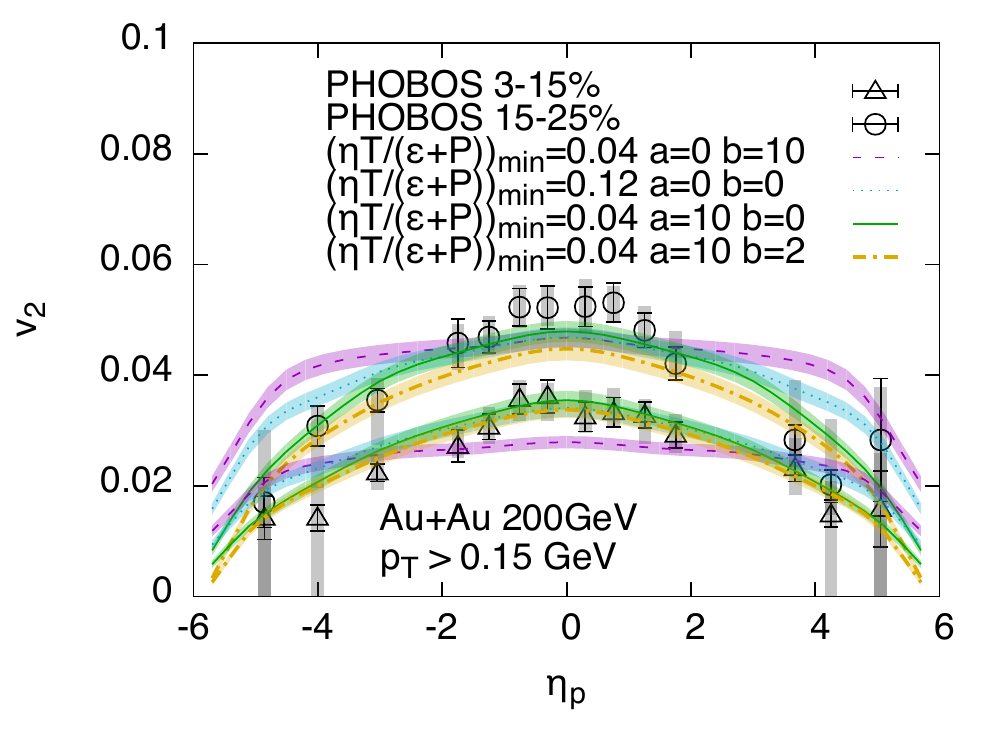}
\caption{(Color online) $v_2$ of charged hadrons as a function of pseudo-rapidity for the four different shear viscosity scenarios compared to experimental data from the PHOBOS collaboration \cite{Back:2004zg,Back:2004mh}. Top: 0-40\% centrality. Bottom 3-15\% and 15-25\% centralities. \label{fig:v2-eta}}
\end{figure}

\begin{figure}[htb]
\includegraphics[width=0.45\textwidth]{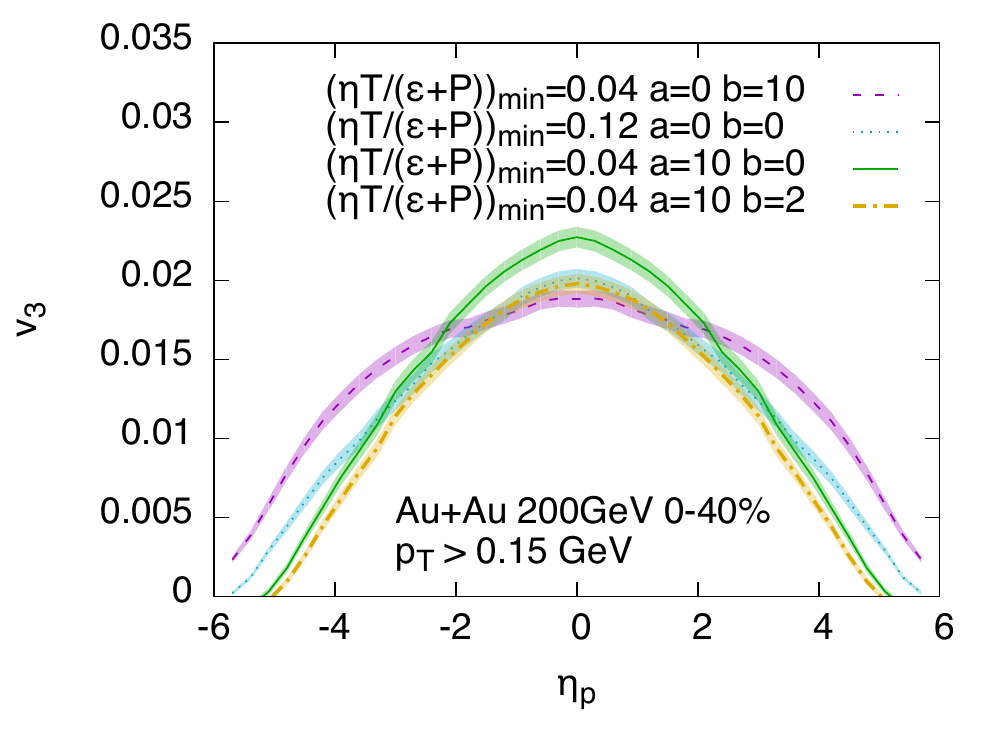}
\caption{(Color online) Prediction for $v_3$ of charged hadrons as a function of pseudo-rapidity for the four scenarios. \label{fig:v3-eta}}
\end{figure}

We show results for the charged hadron $v_2(\eta_p)$ for 0-40\% (top) and 3-15\% and 15-25\% (bottom) central $\sqrt{s}=200\,{\rm GeV}$ collisions and $p_T>0.15\,{\rm GeV}$ in Fig.\,\ref{fig:v2-eta} for the four different scenarios discussed above.\footnote{All results for $v_n(\eta_p)$ were symmetrized around $\eta_p=0$ to increase the statistics.} One can see that different temperature dependencies lead to variations in the $\eta_p$ dependence. 
Because the average temperature decreases with increasing rapidity, a large hadronic shear viscosity causes $v_2(\eta_p)$ to drop more quickly with $|\eta_p|$, while a large QGP viscosity makes the distribution flatter in $\eta_p$. The constant $\eta T/(\varepsilon+P)$ case lies between the two cases. Previous calculations using UrQMD in the low temperature regime, which can be compared to the case of large hadronic viscosity, show a similar trend \cite{Nonaka:2006yn,Werner:2009zza} even though with a smaller effect.

The $v_2$ of charged hadrons as a function of pseudo-rapidity at RHIC has been measured by the PHOBOS \cite{Back:2004mh,Back:2004zg} and STAR \cite{Abelev:2008ae} collaborations. As shown in Fig.\,\ref{fig:v2-eta}, the existing data can already constrain the temperature dependence of $\eta T/(\varepsilon+P)$. Clearly a large hadronic viscosity is favored by the PHOBOS data, while a constant value is hard to reconcile with the experimentally observed decrease of $v_2$ with pseudo-rapidity. Assuming that the initial state is not dramatically different from our model description, a QGP shear viscosity as large as the largest one used in this calculation can be excluded. We note that this scenario predicts a wrong centrality dependence of $v_2$ even at mid-rapidity.
The scenario with large hadronic and moderate QGP shear viscosity is still compatible with most of the data, although slightly below around mid-rapidity in the 15-25\% central case.

In Fig.\,\ref{fig:v3-eta} we show the prediction for the pseudo-rapidity differential triangular flow coefficient $v_3$.
We see a faster drop than for $v_2$ with increasing $|\eta_p|$. 
The measurement of this quantity can serve as a consistency check for the temperature dependence of $\eta/s$ and allow to further constrain the three dimensional fluctuating initial state.

As stated above, the experimentally observed shape of $v_2(\eta_p)$ demands a significant increase of $\eta T/(\varepsilon+P)$ with dropping temperature in the hadronic phase and, at the same time, only a mild or no increase with increasing temperature in the QGP phase. Note that increasing the hadronic viscosity will decrease the magnitude of the elliptic flow coefficient $v_2$ also at $\eta_p=0$, a quantity that is already well described by theory. 
To compensate this effect the minimum value of $\eta/s$ had to be reduced by a factor 3, when compared to the case where an effective viscosity is used, i.e., $\eta/s=0.12$. Hence, the true minimum of the QCD shear viscosity can be significantly smaller than what is predicted when extracting an effective temperature independent $\eta/s$. In our calculations, we find $(\eta/s)_{\textrm{min}} \approx 0.04$ at zero baryon chemical potential, i.e., almost one half of the lower bound conjectured using the AdS/CFT duality \cite{Policastro:2001yc,Kovtun:2004de}.          

\begin{figure}[t]
\includegraphics[width=0.48\textwidth]{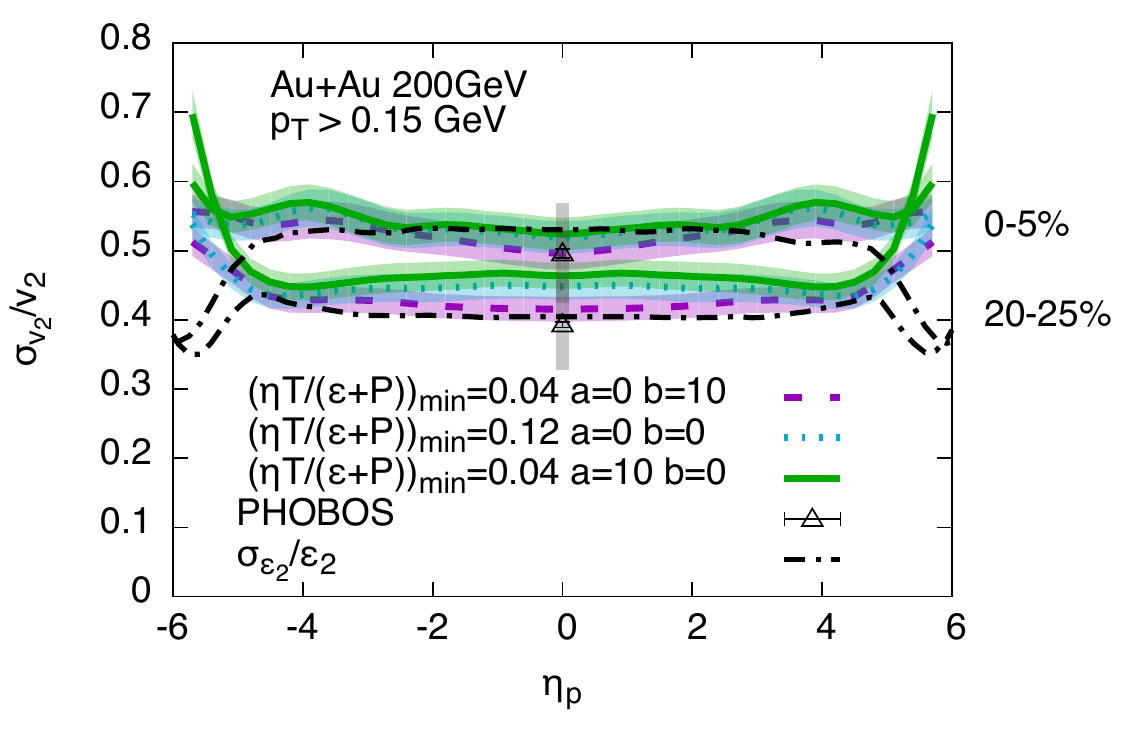}
\caption{(Color online) Variance of the $v_2$ event-by-event distribution for different temperature dependent $\eta T/(\varepsilon+P)$. Dash-dotted lines are the scaled variances of the eccentricity distributions in the initial state. The data points are PHOBOS data \cite{Alver:2007qw} for $N_{\rm part}=214$ ($\approx 20-25\%$) and $296$ ($\approx 0-5\%$). \label{fig:v2-variance-eta}}
\end{figure}

{\bf Rapidity dependent $v_n$ distributions}
At mid-rapidity it was found that the $v_n$ event-by-event distributions \cite{Aad:2013xma} are insensitive to the transport parameters of the medium (when scaled by the mean value) \cite{Niemi:2012aj}. If this is true also at forward rapidities, the distributions could directly be used to constrain the initial state and its fluctuations in three dimensions.
In Fig.\,\ref{fig:v2-variance-eta} we show the (scaled) standard deviation of the $v_2$ distributions vs. pseudo-rapidity $(\sigma_{v_2}/v_2)(\eta_p)$ in the first three scenarios for the shear viscosity temperature dependence. We also compare to the scaled variances of the eccentricity distributions in the initial state. At RHIC this quantity has been measured at mid-rapidity by both PHOBOS \cite{Alver:2007qw} and STAR \cite{Agakishiev:2011eq}.

One can see that 1) at mid-rapidity the scaled variances are compatible with experimental data from PHOBOS \cite{Alver:2007qw}, 2) there is almost no dependence on the pseudo-rapidity in all three cases, 3) final results are close to the initial state results over a wide range in rapidity, and 4) results are only weakly dependent on $\eta T/(\varepsilon+P)$. Thus, the measurement of cumulants of the $v_n$ distributions (or the full distributions) as functions of rapidity will give important information about the 3D initial state and its fluctuations, largely independent of the transport parameters of the medium. In particular it will be interesting to compare predictions for such distributions from more sophisticated initial state models, such as the color glass condensate based IP-Glasma model \cite{Schenke:2012wb,Schenke:2012hg,Gale:2012rq} extended to three dimensions using JIMWLK evolution \cite{Jalilian-Marian:1997jx,Jalilian-Marian:1997gr,Iancu:2000hn,Ferreiro:2001qy,Mueller:2001uk}, because it predicts fluctuation scales that depend on rapidity \cite{Dumitru:2011vk,Schlichting:2014ipa}.

{\bf Conclusions and Outlook}
We have presented results from fully 3+1 dimensional viscous relativistic hydrodynamic simulations including temperature dependent shear and bulk viscosities and using an initial state model that provides three dimensional fluctuating baryon- and entropy densities. 
We have shown that different scenarios for the temperature dependent $\eta T/(\varepsilon+P)$ can lead to significantly different results for the rapidity dependence of elliptic and triangular flow. Comparison with RHIC data provides strong evidence that $\eta T/(\varepsilon+P)$ cannot be constant but must grow with decreasing temperature in the hadronic phase. The case of a strong increase of $\eta T/(\varepsilon+P)$ in the QGP phase ($(\eta T/(\varepsilon+P))(400\,{\rm MeV})\approx 2.4$) is not compatible with the experimental data, while a moderate increase in the QGP ($(\eta T/(\varepsilon+P))(400\,{\rm MeV})\approx 0.5$) cannot be excluded.
We determined the minimum value to be $(\eta/s)_{\textrm{min}} \approx 0.04$, almost one half of the lower bound conjectured using the AdS/CFT duality.
We showed that measurements of $v_3(\eta_p)$ can provide further constraints. 

The event-by-event fluctuations of the flow harmonics are found to be almost insensitive to the transport parameters over a wide range of pseudo-rapidity and thus carry direct information on the fluctuating structure of the produced medium in all spatial dimensions. 
This calls for precise measurements of $v_n$ and their fluctuations over wide ranges in rapidity and at different collision energies at RHIC and LHC. 
They have the potential to eliminate the large theoretical uncertainties in the longitudinal direction and to over-constrain the fluctuating initial state and the temperature dependent transport parameters of QCD matter.

{\bf Acknowledgments} 
AM is supported by the RIKEN Special Postdoctoral Researcher program. GSD and BPS are supported under DOE Contract No. DE-SC0012704. This research used resources of the National Energy Research Scientific Computing Center, which is supported by the Office of Science of the U.S. Department of Energy under Contract No. DE-AC02-05CH11231. BPS acknowledges a DOE Office of Science Early Career Award.

\vspace{-0.5cm}
\bibliography{spires}

\begin{thebibliography}{60}
\expandafter\ifx\csname natexlab\endcsname\relax\def\natexlab#1{#1}\fi
\expandafter\ifx\csname bibnamefont\endcsname\relax
  \def\bibnamefont#1{#1}\fi
\expandafter\ifx\csname bibfnamefont\endcsname\relax
  \def\bibfnamefont#1{#1}\fi
\expandafter\ifx\csname citenamefont\endcsname\relax
  \def\citenamefont#1{#1}\fi
\expandafter\ifx\csname url\endcsname\relax
  \def\url#1{\texttt{#1}}\fi
\expandafter\ifx\csname urlprefix\endcsname\relax\def\urlprefix{URL }\fi
\providecommand{\bibinfo}[2]{#2}
\providecommand{\eprint}[2][]{\url{#2}}

\bibitem[{\citenamefont{Heinz and Snellings}(2013)}]{Heinz:2013th}
\bibinfo{author}{\bibfnamefont{U.}~\bibnamefont{Heinz}} \bibnamefont{and}
  \bibinfo{author}{\bibfnamefont{R.}~\bibnamefont{Snellings}},
  \bibinfo{journal}{Ann. Rev. Nucl. Part. Sci.} \textbf{\bibinfo{volume}{63}},
  \bibinfo{pages}{123} (\bibinfo{year}{2013}), \eprint{1301.2826}.

\bibitem[{\citenamefont{Gale et~al.}(2013{\natexlab{a}})\citenamefont{Gale,
  Jeon, and Schenke}}]{Gale:2013da}
\bibinfo{author}{\bibfnamefont{C.}~\bibnamefont{Gale}},
  \bibinfo{author}{\bibfnamefont{S.}~\bibnamefont{Jeon}}, \bibnamefont{and}
  \bibinfo{author}{\bibfnamefont{B.}~\bibnamefont{Schenke}},
  \bibinfo{journal}{Int. J. of Mod. Phys. A, Vol. 28,}
  \textbf{\bibinfo{volume}{1340011}} (\bibinfo{year}{2013}{\natexlab{a}}),
  \eprint{1301.5893}.

\bibitem[{\citenamefont{de~Souza et~al.}(2015)\citenamefont{de~Souza, Koide,
  and Kodama}}]{deSouza:2015ena}
\bibinfo{author}{\bibfnamefont{R.~D.} \bibnamefont{de~Souza}},
  \bibinfo{author}{\bibfnamefont{T.}~\bibnamefont{Koide}}, \bibnamefont{and}
  \bibinfo{author}{\bibfnamefont{T.}~\bibnamefont{Kodama}}
  (\bibinfo{year}{2015}), \eprint{1506.03863}.

\bibitem[{\citenamefont{Prakash et~al.}(1993)\citenamefont{Prakash, Prakash,
  Venugopalan, and Welke}}]{Prakash:1993bt}
\bibinfo{author}{\bibfnamefont{M.}~\bibnamefont{Prakash}},
  \bibinfo{author}{\bibfnamefont{M.}~\bibnamefont{Prakash}},
  \bibinfo{author}{\bibfnamefont{R.}~\bibnamefont{Venugopalan}},
  \bibnamefont{and} \bibinfo{author}{\bibfnamefont{G.}~\bibnamefont{Welke}},
  \bibinfo{journal}{Phys. Rept.} \textbf{\bibinfo{volume}{227}},
  \bibinfo{pages}{321} (\bibinfo{year}{1993}).

\bibitem[{\citenamefont{Arnold et~al.}(2003)\citenamefont{Arnold, Moore, and
  Yaffe}}]{Arnold:2003zc}
\bibinfo{author}{\bibfnamefont{P.~B.} \bibnamefont{Arnold}},
  \bibinfo{author}{\bibfnamefont{G.~D.} \bibnamefont{Moore}}, \bibnamefont{and}
  \bibinfo{author}{\bibfnamefont{L.~G.} \bibnamefont{Yaffe}},
  \bibinfo{journal}{JHEP} \textbf{\bibinfo{volume}{05}}, \bibinfo{pages}{051}
  (\bibinfo{year}{2003}), \eprint{hep-ph/0302165}.

\bibitem[{\citenamefont{Csernai et~al.}(2006)\citenamefont{Csernai, Kapusta,
  and McLerran}}]{Csernai:2006zz}
\bibinfo{author}{\bibfnamefont{L.~P.} \bibnamefont{Csernai}},
  \bibinfo{author}{\bibfnamefont{J.}~\bibnamefont{Kapusta}}, \bibnamefont{and}
  \bibinfo{author}{\bibfnamefont{L.~D.} \bibnamefont{McLerran}},
  \bibinfo{journal}{Phys. Rev. Lett.} \textbf{\bibinfo{volume}{97}},
  \bibinfo{pages}{152303} (\bibinfo{year}{2006}), \eprint{nucl-th/0604032}.

\bibitem[{\citenamefont{Bilandzic et~al.}(2014)\citenamefont{Bilandzic,
  Christensen, Gulbrandsen, Hansen, and Zhou}}]{Bilandzic:2013kga}
\bibinfo{author}{\bibfnamefont{A.}~\bibnamefont{Bilandzic}},
  \bibinfo{author}{\bibfnamefont{C.~H.} \bibnamefont{Christensen}},
  \bibinfo{author}{\bibfnamefont{K.}~\bibnamefont{Gulbrandsen}},
  \bibinfo{author}{\bibfnamefont{A.}~\bibnamefont{Hansen}}, \bibnamefont{and}
  \bibinfo{author}{\bibfnamefont{Y.}~\bibnamefont{Zhou}},
  \bibinfo{journal}{Phys. Rev.} \textbf{\bibinfo{volume}{C89}},
  \bibinfo{pages}{064904} (\bibinfo{year}{2014}), \eprint{1312.3572}.

\bibitem[{\citenamefont{Aad et~al.}(2014)}]{Aad:2014fla}
\bibinfo{author}{\bibfnamefont{G.}~\bibnamefont{Aad}} \bibnamefont{et~al.}
  (\bibinfo{collaboration}{ATLAS}), \bibinfo{journal}{Phys. Rev.}
  \textbf{\bibinfo{volume}{C90}}, \bibinfo{pages}{024905}
  (\bibinfo{year}{2014}), \eprint{1403.0489}.

\bibitem[{\citenamefont{Niemi et~al.}(2015)\citenamefont{Niemi, Eskola, and
  Paatelainen}}]{Niemi:2015qia}
\bibinfo{author}{\bibfnamefont{H.}~\bibnamefont{Niemi}},
  \bibinfo{author}{\bibfnamefont{K.~J.} \bibnamefont{Eskola}},
  \bibnamefont{and}
  \bibinfo{author}{\bibfnamefont{R.}~\bibnamefont{Paatelainen}}
  (\bibinfo{year}{2015}), \eprint{1505.02677}.

\bibitem[{\citenamefont{Rose et~al.}(2014)\citenamefont{Rose, Paquet, Denicol,
  Luzum, Schenke, Jeon, and Gale}}]{Rose:2014fba}
\bibinfo{author}{\bibfnamefont{J.-B.} \bibnamefont{Rose}},
  \bibinfo{author}{\bibfnamefont{J.-F.} \bibnamefont{Paquet}},
  \bibinfo{author}{\bibfnamefont{G.~S.} \bibnamefont{Denicol}},
  \bibinfo{author}{\bibfnamefont{M.}~\bibnamefont{Luzum}},
  \bibinfo{author}{\bibfnamefont{B.}~\bibnamefont{Schenke}},
  \bibinfo{author}{\bibfnamefont{S.}~\bibnamefont{Jeon}}, \bibnamefont{and}
  \bibinfo{author}{\bibfnamefont{C.}~\bibnamefont{Gale}},
  \bibinfo{journal}{Nucl. Phys.} \textbf{\bibinfo{volume}{A931}},
  \bibinfo{pages}{926} (\bibinfo{year}{2014}), \eprint{1408.0024}.

\bibitem[{\citenamefont{Ryu et~al.}(2015)\citenamefont{Ryu, Paquet, Shen,
  Denicol, Schenke, Jeon, and Gale}}]{Ryu:2015vwa}
\bibinfo{author}{\bibfnamefont{S.}~\bibnamefont{Ryu}},
  \bibinfo{author}{\bibfnamefont{J.~F.} \bibnamefont{Paquet}},
  \bibinfo{author}{\bibfnamefont{C.}~\bibnamefont{Shen}},
  \bibinfo{author}{\bibfnamefont{G.~S.} \bibnamefont{Denicol}},
  \bibinfo{author}{\bibfnamefont{B.}~\bibnamefont{Schenke}},
  \bibinfo{author}{\bibfnamefont{S.}~\bibnamefont{Jeon}}, \bibnamefont{and}
  \bibinfo{author}{\bibfnamefont{C.}~\bibnamefont{Gale}}
  (\bibinfo{year}{2015}), \eprint{1502.01675}.

\bibitem[{\citenamefont{Mishra et~al.}(2008)\citenamefont{Mishra, Mohapatra,
  Saumia, and Srivastava}}]{Mishra:2007tw}
\bibinfo{author}{\bibfnamefont{A.~P.} \bibnamefont{Mishra}},
  \bibinfo{author}{\bibfnamefont{R.~K.} \bibnamefont{Mohapatra}},
  \bibinfo{author}{\bibfnamefont{P.~S.} \bibnamefont{Saumia}},
  \bibnamefont{and} \bibinfo{author}{\bibfnamefont{A.~M.}
  \bibnamefont{Srivastava}}, \bibinfo{journal}{Phys. Rev.}
  \textbf{\bibinfo{volume}{C77}}, \bibinfo{pages}{064902}
  (\bibinfo{year}{2008}), \eprint{0711.1323}.

\bibitem[{\citenamefont{Takahashi et~al.}(2009)}]{Takahashi:2009na}
\bibinfo{author}{\bibfnamefont{J.}~\bibnamefont{Takahashi}}
  \bibnamefont{et~al.}, \bibinfo{journal}{Phys. Rev. Lett.}
  \textbf{\bibinfo{volume}{103}}, \bibinfo{pages}{242301}
  (\bibinfo{year}{2009}).

\bibitem[{\citenamefont{Alver and Roland}(2010)}]{Alver:2010gr}
\bibinfo{author}{\bibfnamefont{B.}~\bibnamefont{Alver}} \bibnamefont{and}
  \bibinfo{author}{\bibfnamefont{G.}~\bibnamefont{Roland}},
  \bibinfo{journal}{Phys. Rev.} \textbf{\bibinfo{volume}{C81}},
  \bibinfo{pages}{054905} (\bibinfo{year}{2010}).

\bibitem[{\citenamefont{Alver et~al.}(2010{\natexlab{a}})\citenamefont{Alver,
  Gombeaud, Luzum, and Ollitrault}}]{Alver:2010dn}
\bibinfo{author}{\bibfnamefont{B.~H.} \bibnamefont{Alver}},
  \bibinfo{author}{\bibfnamefont{C.}~\bibnamefont{Gombeaud}},
  \bibinfo{author}{\bibfnamefont{M.}~\bibnamefont{Luzum}}, \bibnamefont{and}
  \bibinfo{author}{\bibfnamefont{J.-Y.} \bibnamefont{Ollitrault}},
  \bibinfo{journal}{Phys.Rev.} \textbf{\bibinfo{volume}{C82}},
  \bibinfo{pages}{034913} (\bibinfo{year}{2010}{\natexlab{a}}).

\bibitem[{\citenamefont{Holopainen et~al.}(2011)\citenamefont{Holopainen,
  Niemi, and Eskola}}]{Holopainen:2010gz}
\bibinfo{author}{\bibfnamefont{H.}~\bibnamefont{Holopainen}},
  \bibinfo{author}{\bibfnamefont{H.}~\bibnamefont{Niemi}}, \bibnamefont{and}
  \bibinfo{author}{\bibfnamefont{K.~J.} \bibnamefont{Eskola}},
  \bibinfo{journal}{Phys.Rev.} \textbf{\bibinfo{volume}{C83}},
  \bibinfo{pages}{034901} (\bibinfo{year}{2011}).

\bibitem[{\citenamefont{Schenke
  et~al.}(2011{\natexlab{a}})\citenamefont{Schenke, Jeon, and
  Gale}}]{Schenke:2010rr}
\bibinfo{author}{\bibfnamefont{B.}~\bibnamefont{Schenke}},
  \bibinfo{author}{\bibfnamefont{S.}~\bibnamefont{Jeon}}, \bibnamefont{and}
  \bibinfo{author}{\bibfnamefont{C.}~\bibnamefont{Gale}},
  \bibinfo{journal}{Phys. Rev. Lett.} \textbf{\bibinfo{volume}{106}},
  \bibinfo{pages}{042301} (\bibinfo{year}{2011}{\natexlab{a}}).

\bibitem[{\citenamefont{Qiu et~al.}(2012)\citenamefont{Qiu, Shen, and
  Heinz}}]{Qiu:2011hf}
\bibinfo{author}{\bibfnamefont{Z.}~\bibnamefont{Qiu}},
  \bibinfo{author}{\bibfnamefont{C.}~\bibnamefont{Shen}}, \bibnamefont{and}
  \bibinfo{author}{\bibfnamefont{U.~W.} \bibnamefont{Heinz}},
  \bibinfo{journal}{Phys. Lett.} \textbf{\bibinfo{volume}{B707}},
  \bibinfo{pages}{151} (\bibinfo{year}{2012}).

\bibitem[{\citenamefont{Gale et~al.}(2013{\natexlab{b}})\citenamefont{Gale,
  Jeon, Schenke, Tribedy, and Venugopalan}}]{Gale:2012rq}
\bibinfo{author}{\bibfnamefont{C.}~\bibnamefont{Gale}},
  \bibinfo{author}{\bibfnamefont{S.}~\bibnamefont{Jeon}},
  \bibinfo{author}{\bibfnamefont{B.}~\bibnamefont{Schenke}},
  \bibinfo{author}{\bibfnamefont{P.}~\bibnamefont{Tribedy}}, \bibnamefont{and}
  \bibinfo{author}{\bibfnamefont{R.}~\bibnamefont{Venugopalan}},
  \bibinfo{journal}{Phys.Rev.Lett.} \textbf{\bibinfo{volume}{110}},
  \bibinfo{pages}{012302} (\bibinfo{year}{2013}{\natexlab{b}}).

\bibitem[{\citenamefont{Pang et~al.}(2015)\citenamefont{Pang, Qin, Roy, Wang,
  and Ma}}]{Pang:2014pxa}
\bibinfo{author}{\bibfnamefont{L.-G.} \bibnamefont{Pang}},
  \bibinfo{author}{\bibfnamefont{G.-Y.} \bibnamefont{Qin}},
  \bibinfo{author}{\bibfnamefont{V.}~\bibnamefont{Roy}},
  \bibinfo{author}{\bibfnamefont{X.-N.} \bibnamefont{Wang}}, \bibnamefont{and}
  \bibinfo{author}{\bibfnamefont{G.-L.} \bibnamefont{Ma}},
  \bibinfo{journal}{Phys. Rev.} \textbf{\bibinfo{volume}{C91}},
  \bibinfo{pages}{044904} (\bibinfo{year}{2015}).

\bibitem[{\citenamefont{Bozek}(2012)}]{Bozek:2011ua}
\bibinfo{author}{\bibfnamefont{P.}~\bibnamefont{Bozek}},
  \bibinfo{journal}{Phys.Rev.} \textbf{\bibinfo{volume}{C85}},
  \bibinfo{pages}{034901} (\bibinfo{year}{2012}).

\bibitem[{\citenamefont{Molnar et~al.}(2014)\citenamefont{Molnar, Holopainen,
  Huovinen, and Niemi}}]{Molnar:2014zha}
\bibinfo{author}{\bibfnamefont{E.}~\bibnamefont{Molnar}},
  \bibinfo{author}{\bibfnamefont{H.}~\bibnamefont{Holopainen}},
  \bibinfo{author}{\bibfnamefont{P.}~\bibnamefont{Huovinen}}, \bibnamefont{and}
  \bibinfo{author}{\bibfnamefont{H.}~\bibnamefont{Niemi}},
  \bibinfo{journal}{Phys. Rev.} \textbf{\bibinfo{volume}{C90}},
  \bibinfo{pages}{044904} (\bibinfo{year}{2014}), \eprint{1407.8152}.

\bibitem[{\citenamefont{Karpenko et~al.}(2015)\citenamefont{Karpenko, Huovinen,
  Petersen, and Bleicher}}]{Karpenko:2015xea}
\bibinfo{author}{\bibfnamefont{I.~A.} \bibnamefont{Karpenko}},
  \bibinfo{author}{\bibfnamefont{P.}~\bibnamefont{Huovinen}},
  \bibinfo{author}{\bibfnamefont{H.}~\bibnamefont{Petersen}}, \bibnamefont{and}
  \bibinfo{author}{\bibfnamefont{M.}~\bibnamefont{Bleicher}},
  \bibinfo{journal}{Phys. Rev.} \textbf{\bibinfo{volume}{C91}},
  \bibinfo{pages}{064901} (\bibinfo{year}{2015}).

\bibitem[{\citenamefont{Nonaka and Bass}(2007)}]{Nonaka:2006yn}
\bibinfo{author}{\bibfnamefont{C.}~\bibnamefont{Nonaka}} \bibnamefont{and}
  \bibinfo{author}{\bibfnamefont{S.~A.} \bibnamefont{Bass}},
  \bibinfo{journal}{Phys. Rev.} \textbf{\bibinfo{volume}{C75}},
  \bibinfo{pages}{014902} (\bibinfo{year}{2007}).

\bibitem[{\citenamefont{Miller et~al.}(2007)\citenamefont{Miller, Reygers,
  Sanders, and Steinberg}}]{Miller:2007ri}
\bibinfo{author}{\bibfnamefont{M.~L.} \bibnamefont{Miller}},
  \bibinfo{author}{\bibfnamefont{K.}~\bibnamefont{Reygers}},
  \bibinfo{author}{\bibfnamefont{S.~J.} \bibnamefont{Sanders}},
  \bibnamefont{and}
  \bibinfo{author}{\bibfnamefont{P.}~\bibnamefont{Steinberg}},
  \bibinfo{journal}{Ann. Rev. Nucl. Part. Sci.} \textbf{\bibinfo{volume}{57}},
  \bibinfo{pages}{205} (\bibinfo{year}{2007}).

\bibitem[{\citenamefont{Gao et~al.}(2014)\citenamefont{Gao, Guzzi, Huston, Lai,
  Li, Nadolsky, Pumplin, Stump, and Yuan}}]{Gao:2013xoa}
\bibinfo{author}{\bibfnamefont{J.}~\bibnamefont{Gao}},
  \bibinfo{author}{\bibfnamefont{M.}~\bibnamefont{Guzzi}},
  \bibinfo{author}{\bibfnamefont{J.}~\bibnamefont{Huston}},
  \bibinfo{author}{\bibfnamefont{H.-L.} \bibnamefont{Lai}},
  \bibinfo{author}{\bibfnamefont{Z.}~\bibnamefont{Li}},
  \bibinfo{author}{\bibfnamefont{P.}~\bibnamefont{Nadolsky}},
  \bibinfo{author}{\bibfnamefont{J.}~\bibnamefont{Pumplin}},
  \bibinfo{author}{\bibfnamefont{D.}~\bibnamefont{Stump}}, \bibnamefont{and}
  \bibinfo{author}{\bibfnamefont{C.~P.} \bibnamefont{Yuan}},
  \bibinfo{journal}{Phys. Rev.} \textbf{\bibinfo{volume}{D89}},
  \bibinfo{pages}{033009} (\bibinfo{year}{2014}), \eprint{1302.6246}.

\bibitem[{\citenamefont{Eskola et~al.}(2009)\citenamefont{Eskola, Paukkunen,
  and Salgado}}]{Eskola:2009uj}
\bibinfo{author}{\bibfnamefont{K.~J.} \bibnamefont{Eskola}},
  \bibinfo{author}{\bibfnamefont{H.}~\bibnamefont{Paukkunen}},
  \bibnamefont{and} \bibinfo{author}{\bibfnamefont{C.~A.}
  \bibnamefont{Salgado}}, \bibinfo{journal}{JHEP}
  \textbf{\bibinfo{volume}{04}}, \bibinfo{pages}{065} (\bibinfo{year}{2009}),
  \eprint{0902.4154}.

\bibitem[{\citenamefont{Buckley et~al.}(2015)\citenamefont{Buckley, Ferrando,
  Lloyd, Nordstrom, Page, Rufenacht, Schonherr, and Watt}}]{Buckley:2014ana}
\bibinfo{author}{\bibfnamefont{A.}~\bibnamefont{Buckley}},
  \bibinfo{author}{\bibfnamefont{J.}~\bibnamefont{Ferrando}},
  \bibinfo{author}{\bibfnamefont{S.}~\bibnamefont{Lloyd}},
  \bibinfo{author}{\bibfnamefont{K.}~\bibnamefont{Nordstrom}},
  \bibinfo{author}{\bibfnamefont{B.}~\bibnamefont{Page}},
  \bibinfo{author}{\bibfnamefont{M.}~\bibnamefont{Rufenacht}},
  \bibinfo{author}{\bibfnamefont{M.}~\bibnamefont{Schonherr}},
  \bibnamefont{and} \bibinfo{author}{\bibfnamefont{G.}~\bibnamefont{Watt}},
  \bibinfo{journal}{Eur. Phys. J.} \textbf{\bibinfo{volume}{C75}},
  \bibinfo{pages}{132} (\bibinfo{year}{2015}), \eprint{1412.7420}.

\bibitem[{\citenamefont{Bialas and Bzdak}(2007)}]{Bialas:2006qf}
\bibinfo{author}{\bibfnamefont{A.}~\bibnamefont{Bialas}} \bibnamefont{and}
  \bibinfo{author}{\bibfnamefont{A.}~\bibnamefont{Bzdak}},
  \bibinfo{journal}{Acta Phys. Polon.} \textbf{\bibinfo{volume}{B38}},
  \bibinfo{pages}{159} (\bibinfo{year}{2007}), \eprint{hep-ph/0612038}.

\bibitem[{\citenamefont{Broniowski et~al.}(2009)\citenamefont{Broniowski,
  Rybczynski, and Bozek}}]{Broniowski:2007nz}
\bibinfo{author}{\bibfnamefont{W.}~\bibnamefont{Broniowski}},
  \bibinfo{author}{\bibfnamefont{M.}~\bibnamefont{Rybczynski}},
  \bibnamefont{and} \bibinfo{author}{\bibfnamefont{P.}~\bibnamefont{Bozek}},
  \bibinfo{journal}{Comput. Phys. Commun.} \textbf{\bibinfo{volume}{180}},
  \bibinfo{pages}{69} (\bibinfo{year}{2009}).

\bibitem[{\citenamefont{Jeon and Kapusta}(1997)}]{Jeon:1997bp}
\bibinfo{author}{\bibfnamefont{S.}~\bibnamefont{Jeon}} \bibnamefont{and}
  \bibinfo{author}{\bibfnamefont{J.~I.} \bibnamefont{Kapusta}},
  \bibinfo{journal}{Phys. Rev.} \textbf{\bibinfo{volume}{C56}},
  \bibinfo{pages}{468} (\bibinfo{year}{1997}).

\bibitem[{\citenamefont{Monnai and Schenke}(2015)}]{Monnai:2015sca}
\bibinfo{author}{\bibfnamefont{A.}~\bibnamefont{Monnai}} \bibnamefont{and}
  \bibinfo{author}{\bibfnamefont{B.}~\bibnamefont{Schenke}}
  (\bibinfo{year}{2015}), \eprint{1509.04103}.

\bibitem[{\citenamefont{Schenke et~al.}(2010)\citenamefont{Schenke, Jeon, and
  Gale}}]{Schenke:2010nt}
\bibinfo{author}{\bibfnamefont{B.}~\bibnamefont{Schenke}},
  \bibinfo{author}{\bibfnamefont{S.}~\bibnamefont{Jeon}}, \bibnamefont{and}
  \bibinfo{author}{\bibfnamefont{C.}~\bibnamefont{Gale}},
  \bibinfo{journal}{Phys. Rev.} \textbf{\bibinfo{volume}{C82}},
  \bibinfo{pages}{014903} (\bibinfo{year}{2010}).

\bibitem[{\citenamefont{Schenke
  et~al.}(2011{\natexlab{b}})\citenamefont{Schenke, Jeon, and
  Gale}}]{Schenke:2011bn}
\bibinfo{author}{\bibfnamefont{B.}~\bibnamefont{Schenke}},
  \bibinfo{author}{\bibfnamefont{S.}~\bibnamefont{Jeon}}, \bibnamefont{and}
  \bibinfo{author}{\bibfnamefont{C.}~\bibnamefont{Gale}},
  \bibinfo{journal}{Phys. Rev.} \textbf{\bibinfo{volume}{C85}},
  \bibinfo{pages}{024901} (\bibinfo{year}{2011}{\natexlab{b}}).

\bibitem[{\citenamefont{Borsanyi et~al.}(2014)\citenamefont{Borsanyi, Fodor,
  Hoelbling, Katz, Krieg, and Szabo}}]{Borsanyi:2013bia}
\bibinfo{author}{\bibfnamefont{S.}~\bibnamefont{Borsanyi}},
  \bibinfo{author}{\bibfnamefont{Z.}~\bibnamefont{Fodor}},
  \bibinfo{author}{\bibfnamefont{C.}~\bibnamefont{Hoelbling}},
  \bibinfo{author}{\bibfnamefont{S.~D.} \bibnamefont{Katz}},
  \bibinfo{author}{\bibfnamefont{S.}~\bibnamefont{Krieg}}, \bibnamefont{and}
  \bibinfo{author}{\bibfnamefont{K.~K.} \bibnamefont{Szabo}},
  \bibinfo{journal}{Phys. Lett.} \textbf{\bibinfo{volume}{B730}},
  \bibinfo{pages}{99} (\bibinfo{year}{2014}).

\bibitem[{\citenamefont{Borsanyi et~al.}(2012)\citenamefont{Borsanyi, Fodor,
  Katz, Krieg, Ratti, and Szabo}}]{Borsanyi:2011sw}
\bibinfo{author}{\bibfnamefont{S.}~\bibnamefont{Borsanyi}},
  \bibinfo{author}{\bibfnamefont{Z.}~\bibnamefont{Fodor}},
  \bibinfo{author}{\bibfnamefont{S.~D.} \bibnamefont{Katz}},
  \bibinfo{author}{\bibfnamefont{S.}~\bibnamefont{Krieg}},
  \bibinfo{author}{\bibfnamefont{C.}~\bibnamefont{Ratti}}, \bibnamefont{and}
  \bibinfo{author}{\bibfnamefont{K.}~\bibnamefont{Szabo}},
  \bibinfo{journal}{JHEP} \textbf{\bibinfo{volume}{01}}, \bibinfo{pages}{138}
  (\bibinfo{year}{2012}).

\bibitem[{\citenamefont{Niemi et~al.}(2011)\citenamefont{Niemi, Denicol,
  Huovinen, Molnar, and Rischke}}]{Niemi:2011ix}
\bibinfo{author}{\bibfnamefont{H.}~\bibnamefont{Niemi}},
  \bibinfo{author}{\bibfnamefont{G.~S.} \bibnamefont{Denicol}},
  \bibinfo{author}{\bibfnamefont{P.}~\bibnamefont{Huovinen}},
  \bibinfo{author}{\bibfnamefont{E.}~\bibnamefont{Molnar}}, \bibnamefont{and}
  \bibinfo{author}{\bibfnamefont{D.~H.} \bibnamefont{Rischke}},
  \bibinfo{journal}{Phys.Rev.Lett.} \textbf{\bibinfo{volume}{106}},
  \bibinfo{pages}{212302} (\bibinfo{year}{2011}).

\bibitem[{\citenamefont{Niemi et~al.}(2012)\citenamefont{Niemi, Denicol,
  Huovinen, Molnar, and Rischke}}]{Niemi:2012ry}
\bibinfo{author}{\bibfnamefont{H.}~\bibnamefont{Niemi}},
  \bibinfo{author}{\bibfnamefont{G.}~\bibnamefont{Denicol}},
  \bibinfo{author}{\bibfnamefont{P.}~\bibnamefont{Huovinen}},
  \bibinfo{author}{\bibfnamefont{E.}~\bibnamefont{Molnar}}, \bibnamefont{and}
  \bibinfo{author}{\bibfnamefont{D.}~\bibnamefont{Rischke}},
  \bibinfo{journal}{Phys.Rev.} \textbf{\bibinfo{volume}{C86}},
  \bibinfo{pages}{014909} (\bibinfo{year}{2012}).

\bibitem[{\citenamefont{Liao and Koch}(2010)}]{Liao:2009gb}
\bibinfo{author}{\bibfnamefont{J.}~\bibnamefont{Liao}} \bibnamefont{and}
  \bibinfo{author}{\bibfnamefont{V.}~\bibnamefont{Koch}},
  \bibinfo{journal}{Phys. Rev.} \textbf{\bibinfo{volume}{C81}},
  \bibinfo{pages}{014902} (\bibinfo{year}{2010}).

\bibitem[{\citenamefont{Cleymans et~al.}(2006)\citenamefont{Cleymans, Oeschler,
  Redlich, and Wheaton}}]{Cleymans:2005xv}
\bibinfo{author}{\bibfnamefont{J.}~\bibnamefont{Cleymans}},
  \bibinfo{author}{\bibfnamefont{H.}~\bibnamefont{Oeschler}},
  \bibinfo{author}{\bibfnamefont{K.}~\bibnamefont{Redlich}}, \bibnamefont{and}
  \bibinfo{author}{\bibfnamefont{S.}~\bibnamefont{Wheaton}},
  \bibinfo{journal}{Phys. Rev.} \textbf{\bibinfo{volume}{C73}},
  \bibinfo{pages}{034905} (\bibinfo{year}{2006}), \eprint{hep-ph/0511094}.

\bibitem[{\citenamefont{Alver et~al.}(2011)}]{Alver:2010ck}
\bibinfo{author}{\bibfnamefont{B.}~\bibnamefont{Alver}} \bibnamefont{et~al.}
  (\bibinfo{collaboration}{PHOBOS}), \bibinfo{journal}{Phys. Rev.}
  \textbf{\bibinfo{volume}{C83}}, \bibinfo{pages}{024913}
  (\bibinfo{year}{2011}), \eprint{1011.1940}.

\bibitem[{\citenamefont{Werner et~al.}(2009)\citenamefont{Werner, Hirano,
  Karpenko, Pierog, Porteboeuf, Bleicher, and Haussler}}]{Werner:2009zza}
\bibinfo{author}{\bibfnamefont{K.}~\bibnamefont{Werner}},
  \bibinfo{author}{\bibfnamefont{T.}~\bibnamefont{Hirano}},
  \bibinfo{author}{\bibfnamefont{I.}~\bibnamefont{Karpenko}},
  \bibinfo{author}{\bibfnamefont{T.}~\bibnamefont{Pierog}},
  \bibinfo{author}{\bibfnamefont{S.}~\bibnamefont{Porteboeuf}},
  \bibinfo{author}{\bibfnamefont{M.}~\bibnamefont{Bleicher}}, \bibnamefont{and}
  \bibinfo{author}{\bibfnamefont{S.}~\bibnamefont{Haussler}},
  \bibinfo{journal}{J. Phys.} \textbf{\bibinfo{volume}{G36}},
  \bibinfo{pages}{064030} (\bibinfo{year}{2009}), \eprint{0907.5529}.

\bibitem[{\citenamefont{Back et~al.}(2005{\natexlab{a}})}]{Back:2004mh}
\bibinfo{author}{\bibfnamefont{B.~B.} \bibnamefont{Back}} \bibnamefont{et~al.}
  (\bibinfo{collaboration}{PHOBOS}), \bibinfo{journal}{Phys. Rev.}
  \textbf{\bibinfo{volume}{C72}}, \bibinfo{pages}{051901}
  (\bibinfo{year}{2005}{\natexlab{a}}).

\bibitem[{\citenamefont{Back et~al.}(2005{\natexlab{b}})}]{Back:2004zg}
\bibinfo{author}{\bibfnamefont{B.~B.} \bibnamefont{Back}} \bibnamefont{et~al.}
  (\bibinfo{collaboration}{PHOBOS}), \bibinfo{journal}{Phys. Rev. Lett.}
  \textbf{\bibinfo{volume}{94}}, \bibinfo{pages}{122303}
  (\bibinfo{year}{2005}{\natexlab{b}}), \eprint{nucl-ex/0406021}.

\bibitem[{\citenamefont{Abelev et~al.}(2008)}]{Abelev:2008ae}
\bibinfo{author}{\bibfnamefont{B.~I.} \bibnamefont{Abelev}}
  \bibnamefont{et~al.} (\bibinfo{collaboration}{STAR}), \bibinfo{journal}{Phys.
  Rev.} \textbf{\bibinfo{volume}{C77}}, \bibinfo{pages}{054901}
  (\bibinfo{year}{2008}), \eprint{0801.3466}.

\bibitem[{\citenamefont{Policastro et~al.}(2001)\citenamefont{Policastro, Son,
  and Starinets}}]{Policastro:2001yc}
\bibinfo{author}{\bibfnamefont{G.}~\bibnamefont{Policastro}},
  \bibinfo{author}{\bibfnamefont{D.~T.} \bibnamefont{Son}}, \bibnamefont{and}
  \bibinfo{author}{\bibfnamefont{A.~O.} \bibnamefont{Starinets}},
  \bibinfo{journal}{Phys. Rev. Lett.} \textbf{\bibinfo{volume}{87}},
  \bibinfo{pages}{081601} (\bibinfo{year}{2001}), \eprint{hep-th/0104066}.

\bibitem[{\citenamefont{Kovtun et~al.}(2005)}]{Kovtun:2004de}
\bibinfo{author}{\bibfnamefont{P.}~\bibnamefont{Kovtun}} \bibnamefont{et~al.},
  \bibinfo{journal}{Phys. Rev. Lett.} \textbf{\bibinfo{volume}{94}},
  \bibinfo{pages}{111601} (\bibinfo{year}{2005}).

\bibitem[{\citenamefont{Aad et~al.}(2013)}]{Aad:2013xma}
\bibinfo{author}{\bibfnamefont{G.}~\bibnamefont{Aad}} \bibnamefont{et~al.}
  (\bibinfo{collaboration}{ATLAS}), \bibinfo{journal}{JHEP}
  \textbf{\bibinfo{volume}{11}}, \bibinfo{pages}{183} (\bibinfo{year}{2013}),
  \eprint{1305.2942}.

\bibitem[{\citenamefont{Niemi et~al.}(2013)\citenamefont{Niemi, Denicol,
  Holopainen, and Huovinen}}]{Niemi:2012aj}
\bibinfo{author}{\bibfnamefont{H.}~\bibnamefont{Niemi}},
  \bibinfo{author}{\bibfnamefont{G.~S.} \bibnamefont{Denicol}},
  \bibinfo{author}{\bibfnamefont{H.}~\bibnamefont{Holopainen}},
  \bibnamefont{and} \bibinfo{author}{\bibfnamefont{P.}~\bibnamefont{Huovinen}},
  \bibinfo{journal}{Phys. Rev.} \textbf{\bibinfo{volume}{C87}},
  \bibinfo{pages}{054901} (\bibinfo{year}{2013}), \eprint{1212.1008}.

\bibitem[{\citenamefont{Alver et~al.}(2010{\natexlab{b}})}]{Alver:2007qw}
\bibinfo{author}{\bibfnamefont{B.}~\bibnamefont{Alver}} \bibnamefont{et~al.}
  (\bibinfo{collaboration}{PHOBOS}), \bibinfo{journal}{Phys. Rev. Lett.}
  \textbf{\bibinfo{volume}{104}}, \bibinfo{pages}{142301}
  (\bibinfo{year}{2010}{\natexlab{b}}), \eprint{nucl-ex/0702036}.

\bibitem[{\citenamefont{Agakishiev et~al.}(2012)}]{Agakishiev:2011eq}
\bibinfo{author}{\bibfnamefont{G.}~\bibnamefont{Agakishiev}}
  \bibnamefont{et~al.} (\bibinfo{collaboration}{STAR}), \bibinfo{journal}{Phys.
  Rev.} \textbf{\bibinfo{volume}{C86}}, \bibinfo{pages}{014904}
  (\bibinfo{year}{2012}), \eprint{1111.5637}.

\bibitem[{\citenamefont{Schenke
  et~al.}(2012{\natexlab{a}})\citenamefont{Schenke, Tribedy, and
  Venugopalan}}]{Schenke:2012wb}
\bibinfo{author}{\bibfnamefont{B.}~\bibnamefont{Schenke}},
  \bibinfo{author}{\bibfnamefont{P.}~\bibnamefont{Tribedy}}, \bibnamefont{and}
  \bibinfo{author}{\bibfnamefont{R.}~\bibnamefont{Venugopalan}},
  \bibinfo{journal}{Phys. Rev. Lett.} \textbf{\bibinfo{volume}{108}},
  \bibinfo{pages}{252301} (\bibinfo{year}{2012}{\natexlab{a}}).

\bibitem[{\citenamefont{Schenke
  et~al.}(2012{\natexlab{b}})\citenamefont{Schenke, Tribedy, and
  Venugopalan}}]{Schenke:2012hg}
\bibinfo{author}{\bibfnamefont{B.}~\bibnamefont{Schenke}},
  \bibinfo{author}{\bibfnamefont{P.}~\bibnamefont{Tribedy}}, \bibnamefont{and}
  \bibinfo{author}{\bibfnamefont{R.}~\bibnamefont{Venugopalan}},
  \bibinfo{journal}{Phys. Rev.} \textbf{\bibinfo{volume}{C86}},
  \bibinfo{pages}{034908} (\bibinfo{year}{2012}{\natexlab{b}}).

\bibitem[{\citenamefont{Jalilian-Marian
  et~al.}(1997)\citenamefont{Jalilian-Marian, Kovner, Leonidov, and
  Weigert}}]{Jalilian-Marian:1997jx}
\bibinfo{author}{\bibfnamefont{J.}~\bibnamefont{Jalilian-Marian}},
  \bibinfo{author}{\bibfnamefont{A.}~\bibnamefont{Kovner}},
  \bibinfo{author}{\bibfnamefont{A.}~\bibnamefont{Leonidov}}, \bibnamefont{and}
  \bibinfo{author}{\bibfnamefont{H.}~\bibnamefont{Weigert}},
  \bibinfo{journal}{Nucl. Phys.} \textbf{\bibinfo{volume}{B504}},
  \bibinfo{pages}{415} (\bibinfo{year}{1997}), \eprint{hep-ph/9701284}.

\bibitem[{\citenamefont{Jalilian-Marian
  et~al.}(1999)\citenamefont{Jalilian-Marian, Kovner, Leonidov, and
  Weigert}}]{Jalilian-Marian:1997gr}
\bibinfo{author}{\bibfnamefont{J.}~\bibnamefont{Jalilian-Marian}},
  \bibinfo{author}{\bibfnamefont{A.}~\bibnamefont{Kovner}},
  \bibinfo{author}{\bibfnamefont{A.}~\bibnamefont{Leonidov}}, \bibnamefont{and}
  \bibinfo{author}{\bibfnamefont{H.}~\bibnamefont{Weigert}},
  \bibinfo{journal}{Phys. Rev.} \textbf{\bibinfo{volume}{D59}},
  \bibinfo{pages}{014014} (\bibinfo{year}{1999}), \eprint{hep-ph/9706377}.

\bibitem[{\citenamefont{Iancu et~al.}(2001)\citenamefont{Iancu, Leonidov, and
  McLerran}}]{Iancu:2000hn}
\bibinfo{author}{\bibfnamefont{E.}~\bibnamefont{Iancu}},
  \bibinfo{author}{\bibfnamefont{A.}~\bibnamefont{Leonidov}}, \bibnamefont{and}
  \bibinfo{author}{\bibfnamefont{L.~D.} \bibnamefont{McLerran}},
  \bibinfo{journal}{Nucl. Phys.} \textbf{\bibinfo{volume}{A692}},
  \bibinfo{pages}{583} (\bibinfo{year}{2001}), \eprint{hep-ph/0011241}.

\bibitem[{\citenamefont{Ferreiro et~al.}(2002)\citenamefont{Ferreiro, Iancu,
  Leonidov, and McLerran}}]{Ferreiro:2001qy}
\bibinfo{author}{\bibfnamefont{E.}~\bibnamefont{Ferreiro}},
  \bibinfo{author}{\bibfnamefont{E.}~\bibnamefont{Iancu}},
  \bibinfo{author}{\bibfnamefont{A.}~\bibnamefont{Leonidov}}, \bibnamefont{and}
  \bibinfo{author}{\bibfnamefont{L.}~\bibnamefont{McLerran}},
  \bibinfo{journal}{Nucl. Phys.} \textbf{\bibinfo{volume}{A703}},
  \bibinfo{pages}{489} (\bibinfo{year}{2002}), \eprint{hep-ph/0109115}.

\bibitem[{\citenamefont{Mueller}(2001)}]{Mueller:2001uk}
\bibinfo{author}{\bibfnamefont{A.~H.} \bibnamefont{Mueller}},
  \bibinfo{journal}{Phys. Lett.} \textbf{\bibinfo{volume}{B523}},
  \bibinfo{pages}{243} (\bibinfo{year}{2001}), \eprint{hep-ph/0110169}.

\bibitem[{\citenamefont{Dumitru et~al.}(2011)\citenamefont{Dumitru,
  Jalilian-Marian, Lappi, Schenke, and Venugopalan}}]{Dumitru:2011vk}
\bibinfo{author}{\bibfnamefont{A.}~\bibnamefont{Dumitru}},
  \bibinfo{author}{\bibfnamefont{J.}~\bibnamefont{Jalilian-Marian}},
  \bibinfo{author}{\bibfnamefont{T.}~\bibnamefont{Lappi}},
  \bibinfo{author}{\bibfnamefont{B.}~\bibnamefont{Schenke}}, \bibnamefont{and}
  \bibinfo{author}{\bibfnamefont{R.}~\bibnamefont{Venugopalan}},
  \bibinfo{journal}{Phys.Lett.} \textbf{\bibinfo{volume}{B706}},
  \bibinfo{pages}{219} (\bibinfo{year}{2011}).

\bibitem[{\citenamefont{Schlichting and Schenke}(2014)}]{Schlichting:2014ipa}
\bibinfo{author}{\bibfnamefont{S.}~\bibnamefont{Schlichting}} \bibnamefont{and}
  \bibinfo{author}{\bibfnamefont{B.}~\bibnamefont{Schenke}},
  \bibinfo{journal}{Phys.Lett.} \textbf{\bibinfo{volume}{B739}},
  \bibinfo{pages}{313} (\bibinfo{year}{2014}), \eprint{1407.8458}.

\end{thebibliography}

\end{document}